# Atomic force microscopy study of the surface degradation mechanisms of zirconia based ceramics

Sylvain Deville[a], Jérôme Chevalier[a], Gilbert Fantozzi[a], Ramón Torrecillas[b], José F. Bartolomé[c], José S. Moya[c]

[a] National Institute of Applied Science, Materials Department
Associate Research Unit 5510, Bat B.Pascal, 20 av. Albert Einstein
69621 Villeurbanne, France
[b] Instituto Nacional del Carbón
C/ Francisco Pintado Fe, 26, La Corredoira, 33011 Oviedo, SPAIN
[c] Instituto de Ciencia de Materiales de Madrid, Cantoblanco 28049, Madrid, SPAIN

**Abstract**

Atomic force microscopy (AFM) can be used to characterise several aspects of the surface degradation and reinforcement mechanisms of zirconia based ceramics, such as crack propagation, martensitic relief formation, grains pull-out and transformation toughening. AFM can also be used to quantify precisely the transformation and provide reliable parameters for long term degradation prediction. In particular, the tetragonal to monoclinic (t-m) phase transformation of zirconia has been the object of extensive investigations of the last twenty years, and is now recognised as being of martensitic nature. New strong evidences supporting the martensitic nature of the transformation are reported here. These observations, considering their scale and precision, are a new step toward the understanding of the t-m phase transformation of zirconia and related degradation mechanisms.

**Introduction**

Among engineering ceramics, zirconia based ceramics represent a very large range of materials, for which applications are countless, from gas filters to bioceramics for orthopaedic applications. The discovery of the transformation toughening effect in the 70's was followed by an extensive period of deep investigations in order to understand the underlying mechanisms and potential implications and applications of this phenomenon This behaviour is accounted for by the ability of zirconia to transform from its metastable tetragonal phase at ambient temperature to its stable monoclinic phase, under the action of thermal or mechanical stress. The volume increase related to the structural differences is the source of increased crack resistance propagation, partially closing the crack when propagating.

Though very beneficial in terms of mechanical properties, this phase transformation has also proven to be very detrimental in terms of surface degradation. Zirconia might transform at low temperature (even at room temperature), especially under humid conditions, leading to grain pop-up and pull-out, microcracks formations and subsequent surface degradation. This phenomenon, occurring with time at the surface of zirconia is now referred as aging. There is now a regain of interest from a biomedical point of view, since degradation and rupture of hip joints prosthesis related to the transformation have been reported recently.

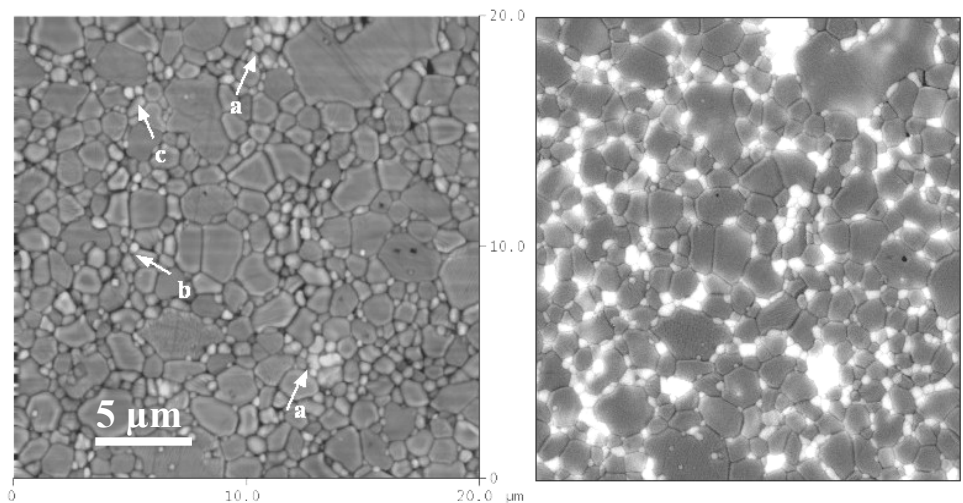

Fig. 1: Observation of the same zone of ZTA by AFM (left, relief contrast) and SEM (right, species contrast) after autoclave thermal ageing treatment at 413 K. The transformation was found to occur either in zirconia agglomerates (a) or isolated grains (b and c). The comparison of AFM and SEM images allowed characterizing the microstructural environment of transformed zirconia grains.

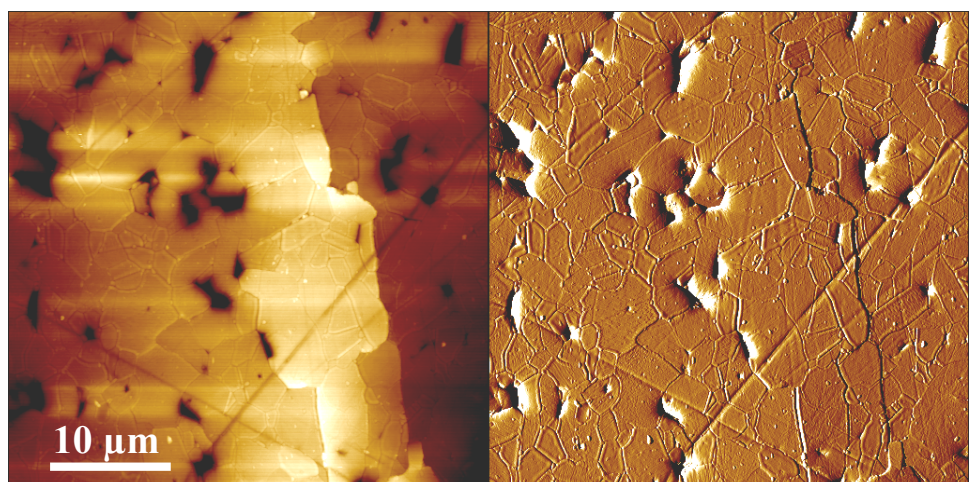

Fig. 2: AFM observation of a propagated crack on a (1.6 vol.% $ZrO_2$)-alumina sample (height image and deflection image). The Vickers indent induced crack path is clearly identified, as well as the crack bridging zones. Alumina grain pull-outs resulting from the polishing process are clearly visible.

From a fundamental point of view, the transformation is now recognised as being of martensitic nature: diffusionless, athermal, involving shape change dominated by shear and occurring at high speed. The transformation has been observed on specimens with direct methods such as interferometry and Normarski interference, or *in situ* on thin foils by transmission electron microscopy. Indirect measurements by thermal differential analysis or X-Ray diffraction provide with informations on the global behavior of the material.

The underlying mechanisms governing the transformation are still a matter of debate. Further progress in the understanding of the phenomenon requires local direct observation at grain-size scale and in real conditions (bulk samples) if possible.

Considering the scale at which surface modifications are occurring, the very low number of studies that might be found so far is not surprising. Quite fortunately, the development of scanning tunnelling microscopy (STM) and atomic force microscopy (AFM) in the last ten years provides the scientific community with very powerful tools to investigate the surface relief at a nanometer scale. Very few studies[1,2] using AFM have been reported so far, and since they were still preliminary, they provide little quantitative informations on the transformation.

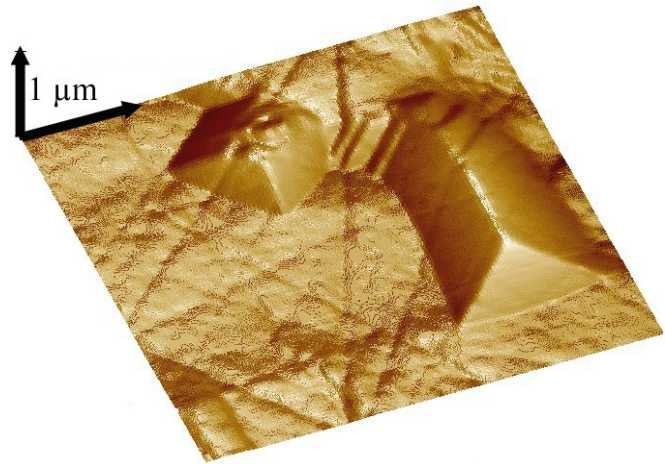

Fig. 3: Self accommodating martensitic variant pairs formed in the surroundings of a propagated crack of a double torsion test Ce-TZP sample.

In the results reported here, AFM has been used to investigate very locally several aspects of the transformation: quantification and prediction of the long term ageing behaviour, microstructural analysis of the ageing behaviour of zirconia toughened alumina, transformation toughening mechanisms analysis, influence of the cubic phase on the ageing behaviour, crack propagation characterization and more importantly martensitic relief observation and characterization.

## Experimental methods

Yttria stabilized samples were processed using yttria-stabilized zirconia powder (3Y-TZP, Tosoh TZ-3YS, Tosoh Corporation, Tokyo, Japan). Green bodies, after uniaxial cold pressing at 10 bars, were isostatically cold pressed at 1300 bars and finally sintered at 1773 K for five hours in air. Zirconia toughened alumina samples were processed by a classical powder mixing processing route. A high purity alumina powder α-$Al_2O_3$ >99.9 wt.% (Condea HPA 0.5, Ceralox division, Arizona, USA) was mixed with 25 wt.% (i.e. 16.7 vol.%) of 3Y-TZP powder. Samples were sintered at 1873 K for two hours in air. Ceria stabilized zirconia ($CeO_2$-TZP) materials were processed by classical powder mixing processing route, using Zirconia Sales Ltd. powders, with uniaxial pressing and sintering at 1813 K for two hours.

Mirror polished samples were cleaned with ethanol and observed by atomic force microscopy (D3100, Digital Instruments Inc.) in contact mode, using oxide sharpened silicon nitride probes with an average scanning speed of 2 µm/s. Autoclave ageing tests were performed in steam autoclave at various temperatures for different times. Scanning electron microscopy observations were carried out with an ESEM FEG (FEI) in backscattered electrons imaging mode on gold coated samples.

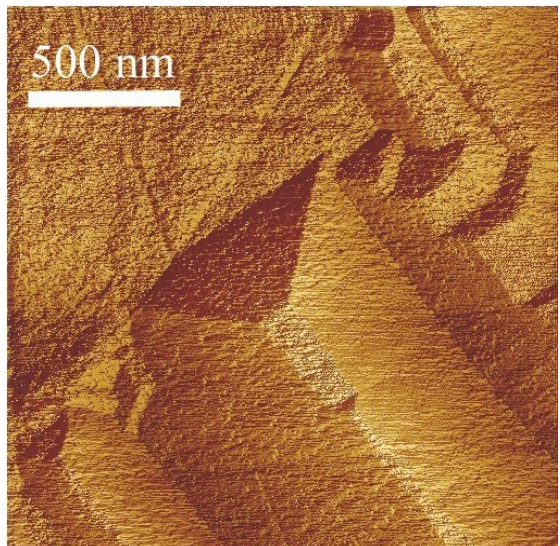

Fig. 4: Observation of thermal martensite self-accommodating variant pairs at the surface of a Ce-TZP sample. The characteristic shape of the relief could be explained by the phenomenological theory of the martensite crystallography.

# Results and discussion

## Measurements and prediction of the long-term ageing behaviour

By following the transformation of a constant zone as a function of the autoclave ageing treatment time, it was possible assessing the nucleation and growth nature of the transformation.

The size of a statistically significant number of monoclinic spots was measured and found to vary linearly with their age, even in the very first stages of growth, where only a fraction of a single grain is transformed. The measured parameters could be incorporated into the prediction models of the transformation[3].

## Microstructural analysis of the ageing behaviour of zirconia toughened alumina

AFM was used to analyse the influence of microstructural parameters in zirconia toughened alumina composites, were alumina is the matrix phase. This kind of materials has been the object of several studies, and factors controlling the transformation have been underlined before[4]. The size effect has been proven to be the most significant factor in controlling the transformation, though other factors such as residual stresses and interface energy between the two phases are also to be considered. The existence of a percolation threshold of the zirconia content, above which transformation is propagating, has been demonstrated recently[5]. However, very little studies are reported concerning the local observation of the transformation and the influence of microstructural parameters. Using a combined analysis with SEM and AFM on the same zone of bulk samples surface (Fig. 1), the influence of zirconia agglomerates on the transformation was analysed. It was shown that either grains within the zirconia agglomerates and small single grains embedded within the matrix were transformable. As previously hypothesised, the transformation of small single grains might be related to the presence of large residual stresses resulting from the difference between thermal expansion coefficients of alumina and zirconia.

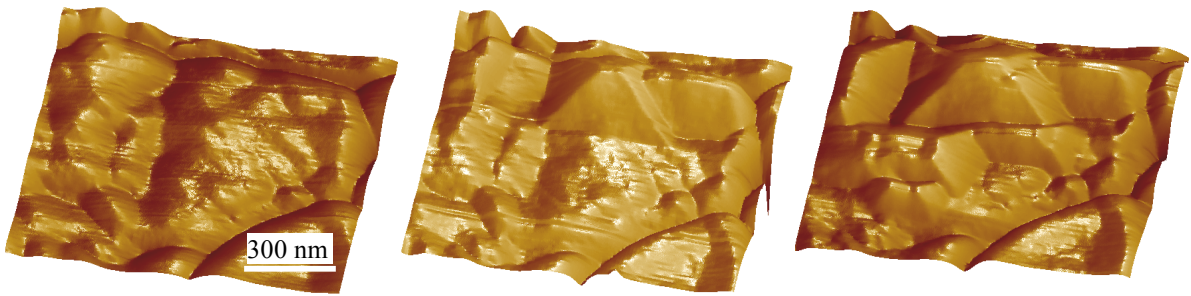

Fig. 5: Progressive transformation of a single grain in 3Y-TZP at 383 K (0, 5 and 10 min). The apparition and growth of the variant pairs is clearly observed. (Vertical scale 35 nm)

**Crack propagation characterization**

AFM was used to investigate the crack propagation path at the surface of double torsion samples and Vickers indent induced cracks. A preliminary thermal etching treatment was performed so as to reveal the grain boundary. Grain boundary grooves and the slight grains relief induced by thermal etching, related to the crystallographic orientation of the grain with respect to the surface were used to study the inter- or intragranular nature of the crack propagation. AFM observations (Fig. 2) were found to be much more precise than scanning electron microscopy. In particular, the observation of crack branching and bridging zones is very clear.

**Transformation toughening mechanisms analysis**

The relief in the surroundings of propagated crack was observed locally on Ce-TZP double torsion samples. It was found possible observing martensitic relief at the surface of transformed grains (Fig. 3). Depending on the stabilizing oxide content, the transformation was found to be very easily stress induced or not.

**Effect of the cubic phase on the ageing behaviour**

3Y-TZP samples sintered at high temperatures (i.e. above 1723 K) were characterised and found to exhibit a dual phase microstructure at ambient temperature with a significant amount of large grains of cubic phase, in agreement with the zirconia-yttria phase diagram. The autoclave ageing stability of the cubic phase was assessed by AFM (Fig. 6). More importantly, the detrimental effect of the cubic phase on the ageing behaviour has been proven, in that the yttrium depletion zones around yttria enriched cubic grains appeared as preferred nucleation sites[6].

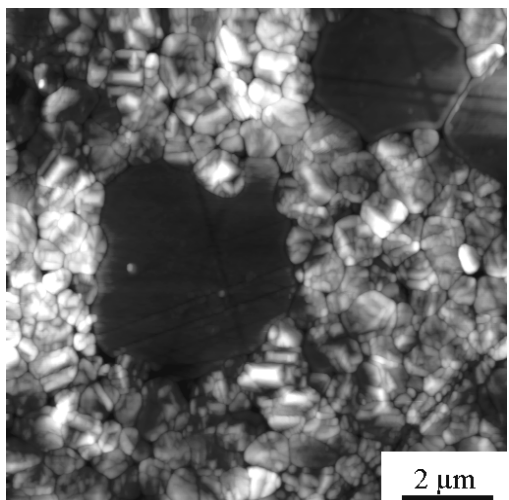

Fig. 6: Surface transformation of a 3Y-TZP sample exhibiting a large amount of cubic phase grains, after 24 hrs of autoclave ageing at 413 K. The large cubic grains are clearly not transformed.

**Martensitic relief observation and characterization**

AFM has been used to investigate very locally the martensitic aspects of the transformation. The main discovery is the observation[7] for the first time of very clear martensitic relief at a nanometer scale, on several types of materials (3Y-TZP, 10Ce-TZP, and $Al_2O_3$-(3Y-TZP)). It was shown that using AFM, it is possible to observe the apparition of self accommodating martensitic variants at the surface of bulks samples, after thermal ageing (Fig. 4). The progressive transformation of each grain, step by step (Fig. 5), is clearly observed. One of the great points using AFM is the possible quantitative measurements of surface relief, e.g. angle relationships between self accommodating variants, etc... Those informations could be related to the underlying crystallography, and the observations reported here fit very well the theoretical predictions of phenomenological theories.

## Conclusions

The purpose of this communication is to show that AFM is an extremely powerful tool to investigate the surface degradation mechanisms of zirconia containing ceramics. AFM is a unique method to provide 3D measurements at a nanometer scale with relative ease of image interpretation. In particular, it is possible imaging nanometer scale martensitic relief, providing very valuable informations for the mechanistic understanding.

## Acknowledgments

Financial support of the European Union under the GROWTH 2000 program, project BIOKER, reference GRD2-2000- 25039.

## References

[1] X. Y. Chen, X. H. Zheng, H. S. Fang, H. Z. Shi, X. F. Wang, H. M. Chen, "The study of martensitic transformation and nanoscale surface relief in zirconia", *Journal of Materials Science Letters*, **21** 415-418 (2002).

[2] H. Tsubakino, Y. Kuroda, M. Niibe, "Surface relief associated with isothermal martensite in zirconia 3 mol% yttria ceramics observed by atomic force microscopy", *Communication of the American Ceramic Society*, **82** [10] 2921-23 (1999).

[3] L. Gremillard, J. Chevalier, S. Deville, T. Epicier, G. Fantozzi, "Modeling the aging kinetics of zirconia ceramics", *Journal of the European Ceramic Society*, (accepted).

[4] S. Deville, J. Chevalier, G. Fantozzi, J. F. Bartolomé, J. Requena, J. S. Moya, R. Torrecillas, L. A. Diaz, "Low temperature ageing of zirconia toughened alumina ceramics and its biomedical implications", *Journal of the European Ceramic Society*, **23** 2975-2982 (2003).

[5] C. Pecharroman, J. F. Bartolomé, J. Requena, J. S. Moya, S. Deville, J. Chevalier, G. Fantozzi, R. Torrecillas, "Percolative mechanisms of aging in zirconia containing ceramics for medical applications", *Advanced Materials*, **15** [6] 507-511 (2003).

[6] J. Chevalier, S. Deville, E. Münch, R. Jullian, F. Lair "Critical effect of cubic phase on aging in 3 mol.% yttria stabilized zirconia ceramics", *Biomaterials* (accepted).

[7] S. Deville, J. Chevalier, "Martensitic relief observation by atomic force microscopy in yttria stabilized zirconia", *Journal of the American Ceramic Society*, **86** [12] 2225-2227(2003).